\documentclass[11pt, oneside, english]{article}
\usepackage{jheppub}

\usepackage{amsmath}
\usepackage{amssymb}
\usepackage{amsfonts}
\usepackage{amsthm}
\usepackage{amsbsy}
\usepackage{array}

\usepackage[T1]{fontenc}
\usepackage[latin9]{inputenc}
\usepackage{url}
\usepackage{graphicx}

\makeatletter
\newenvironment{lyxcode}
	{\par\begin{list}{}{
		\setlength{\rightmargin}{\leftmargin}
		\setlength{\listparindent}{0pt}
		\raggedright
		\setlength{\itemsep}{0pt}
		\setlength{\parsep}{0pt}
		\normalfont\ttfamily}%
	 \item[]}
	{\end{list}}

\usepackage{hyperref}

\makeatother

\usepackage{babel}

\makeatletter
\def\@fpheader{\ }
\makeatother

\title{Scaling the semidefinite program solver SDPB}
\author{Walter Landry$^{1,2}$ and David Simmons-Duffin$^{1}$}
\affiliation{
${}^1$Walter Burke Institute for Theoretical Physics, Caltech, Pasadena, California 91125, USA \\
${}^2$Simons Collaboration on the Nonperturbative Bootstrap
}
\emailAdd{wlandry@caltech.edu}
\emailAdd{dsd@caltech.edu}

\abstract{
We present enhancements to SDPB, an open source, parallelized, arbitrary
precision semidefinite program solver designed for the conformal bootstrap.
The main enhancement is significantly improved performance and scalability
using the Elemental library and MPI. The result is a new version of SDPB that runs on multiple
nodes with hundreds of cores with excellent scaling, making it practical
to solve larger problems. We demonstrate performance on a moderate-size problem in the 3d Ising CFT and a much larger problem in the $O(2)$ Model.
}

\preprint{CALT-TH 2019-038}

\begin{document}
\maketitle

\section{Introduction}

The conformal bootstrap program \cite{Ferrara:1973yt,Polyakov:1974gs} for constraining and solving conformal field theories (CFTs) experienced a renaissance a decade ago with the
introduction of numerical bootstrap methods \cite{Rattazzi:2008pe}. Over the past several years,
these methods have been further developed, generalized, and improved,
leading to many advances in our understanding of CFTs. 
An example application of numerical bootstrap methods is a high-precision determination of
 scaling dimensions and OPE coefficients in the 3d Ising CFT \cite{ElShowk:2012ht,El-Showk:2014dwa,Kos:2014bka,Simmons-Duffin:2015qma,Kos:2016ysd,Komargodski:2016auf,Simmons-Duffin:2016wlq}. See \cite{Poland:2018epd,Chester:2019wfx} for recent reviews of numerical bootstrap methods.

The key ingredients in the numerical bootstrap are crossing symmetry of four-point correlation functions and unitarity. The space of solutions to these constraints can be explored numerically using semidefinite programming \cite{Poland:2011ey,Kos:2013tga,Kos:2014bka,Simmons-Duffin:2015qma}.
In \cite{Simmons-Duffin:2015qma}, one of us (DSD) introduced the semidefinite program solver SDPB, designed for   bootstrap applications. SDPB implements a well-known primal-dual interior-point method for solving semidefinite programs \cite{doi:10.1137/1.9781611970791,alizadeh,citeulike:3473686,Vandenberghe:1995:PPR:213565.213573,SDPA,SDPA2,SDPAGMP}. Its distinguishing features are:
\begin{enumerate}
\item It takes advantage of sparsity patterns in matrices that arise in bootstrap problems;
\item It uses arbitrary-precision arithmetic, making it much more robust (though slower) than machine-precision solvers.\footnote{SDPA-GMP \cite{SDPAGMP} is another arbitrary precision semidefinite program solver, which was used in some bootstrap works before the arrival of SDPB.}
\end{enumerate}

Since its introduction, SDPB 1.0 has been used in approximately 70 works \cite{Kos:2016ysd,
Simmons-Duffin:2016wlq,
Kos:2015mba,
Chester:2015qca,
Beem:2015aoa,
Iliesiu:2015qra,
Shimada:2015gda,
Poland:2015mta,
Lemos:2015awa,
Lin:2015wcg,
Chester:2015lej,
Chester:2016wrc,
Behan:2016dtz,
Nakayama:2016jhq,
Iha:2016ppj,
Komargodski:2016auf,
Nakayama:2016knq,
Echeverri:2016ztu,
Paulos:2016fap,
Paulos:2016but,
Li:2016wdp,
Collier:2016cls,
Pang:2016xno,
Lin:2016gcl,
Bae:2016yna,
Bae:2016jpi,
Lemos:2016xke,
Beem:2016wfs,
Li:2017ddj,
Collier:2017shs,
Cornagliotto:2017dup,
Behan:2017emf,
Nakayama:2017vdd,
Iliesiu:2017nrv,
Dymarsky:2017xzb,
Chang:2017xmr,
Cho:2017fzo,
Dymarsky:2017yzx,
Paulos:2017fhb,
Bae:2017kcl,
Dyer:2017rul,
Chester:2017vdh,
Chang:2017cdx,
Cornagliotto:2017snu,
Agmon:2017xes,
Rong:2017cow,
Baggio:2017mas,
Stergiou:2018gjj,
Poland:2018epd,
Liendo:2018ukf,
Rong:2018okz,
Atanasov:2018kqw,
Behan:2018hfx,
Kousvos:2018rhl,
Bae:2018qym,
Bao:2018dzk,
Cappelli:2018vir,
Gowdigere:2018lxz,
Li:2018lyb,
Karateev:2019pvw,
Afkhami-Jeddi:2019zci,
Go:2019lke,
Stergiou:2019dcv,
Lin:2019kpn,
delaFuente:2019hbl,
Homrich:2019cbt,
Manenti:2019kbl,
Gimenez-Grau:2019hez,
Chester:2019wfx,
Agmon:2019imm,
Bercini:2019vme}. For example, it played an essential role in the recent high-precision determination of CFT data in the 3d Ising model \cite{Kos:2016ysd,Simmons-Duffin:2016wlq}. It has also been used for exploring theories in 2 dimensions \cite{delaFuente:2019hbl,Lin:2015wcg,Cornagliotto:2017dup,Collier:2017shs,Lin:2016gcl}, 3 dimensions \cite{Iliesiu:2017nrv,Dymarsky:2017xzb,Dymarsky:2017yzx,Atanasov:2018kqw,Agmon:2019imm,Kousvos:2018rhl,Li:2018lyb,Stergiou:2019dcv,Agmon:2017xes,Rong:2017cow,Baggio:2017mas,Stergiou:2018gjj,Chester:2017vdh,Bae:2016jpi,Echeverri:2016ztu,Komargodski:2016auf,Nakayama:2016jhq,Chester:2016wrc,Chester:2015qca,Iliesiu:2015qra,Shimada:2015gda,Kos:2016ysd,Simmons-Duffin:2016wlq,Kos:2015mba}, 4 dimensions \cite{Karateev:2019pvw,Rong:2018okz,Poland:2015mta,Lemos:2015awa,Manenti:2019kbl,Cornagliotto:2017snu,Lemos:2016xke,Beem:2016wfs,Li:2017ddj,Nakayama:2016knq,Iha:2016ppj}, 5 dimensions \cite{Chang:2017cdx,Li:2016wdp}, and 6 dimensions \cite{Beem:2015aoa,Chang:2017xmr}, as well as theories in fractional dimensions \cite{Cappelli:2018vir,Gowdigere:2018lxz,Nakayama:2017vdd,Pang:2016xno,Chester:2015lej}, defect theories \cite{Liendo:2018ukf,Gimenez-Grau:2019hez}, and nonlocal theories \cite{Behan:2017emf,Behan:2018hfx}, and for studying the modular bootstrap \cite{Collier:2016cls,Bae:2016yna,Cho:2017fzo,Bae:2017kcl,Dyer:2017rul,Lin:2019kpn,Afkhami-Jeddi:2019zci,Bae:2018qym}, and the $S$-matrix bootstrap \cite{Paulos:2017fhb,Homrich:2019cbt,Paulos:2016fap,Paulos:2016but,Bercini:2019vme}.

Despite the myriad applications of SDPB 1.0, it is easy to find numerical bootstrap problems that are essentially too large for it. A typical crossing symmetry equation can be written in the form
\begin{align}
\label{eq:crossingequation}
\sum_{R,\Delta,\ell} \vec \lambda_{R,\Delta,\ell}^T V^{I}_{R,\Delta,\ell}(z,\bar z) \vec \lambda_{R,\Delta,\ell} &= 0,\qquad I=1,\dots,I_\mathrm{max}.
\end{align}
Here, $R$ runs over types of operators (e.g., global symmetry representations) appearing in the crossing equation, $\Delta$ runs over scaling dimensions, and $\ell$ runs over spins. The $\vec\lambda_{R,\Delta,\ell}$ are vectors of OPE coefficients.  $V^I_{R,\Delta,\ell}(z,\bar z)$ are matrices whose entries are functions of the cross-ratios $z,\bar z$. The size of a bootstrap computation depends roughly on three quantities:\footnote{Several other quantities also affect the complexity of a bootstrap calculation, but most of them can be tied to $\Lambda$. For example, there is a cutoff on spin $\ell \leq \ell_\mathrm{max}$, and the value of this cutoff depends on $\Lambda$. Furthermore, the required precision of the arithmetic increases with $\Lambda$. For simplicity, we use $\Lambda$ as a proxy for these effects.}
\begin{itemize}
\item The number of crossing equations $I_\mathrm{max}$.
\item The dimension of the OPE coefficient vectors $\dim(\vec \lambda_{R,\Delta,\ell})$ (equivalently, the number of rows or columns in the matrices $V^I_{R,\Delta,\ell}$).
\item The number of derivatives $\partial_z^m \partial_{\bar z}^n$ of the crossing equations used in the semidefinite program. We parameterize the number of derivatives by $\Lambda=\mathrm{max}(m + n)$.
\end{itemize}
Another measure of complexity is the number of components $N$ of the linear functional applied to the crossing equation (\ref{eq:crossingequation}), which is essentially a function of $I_\mathrm{max}$ and $\Lambda$.

As an example, in the computation of the 3d Ising model critical dimensions in \cite{Kos:2016ysd}, performed using SDPB 1.0, the number of crossing equations used was $I_\mathrm{max}=5$, the OPE coefficient vectors were either $1$ or $2$ dimensional, the number of derivatives was $\Lambda=43$, and the resulting linear functional had $N=1265$ components. That computation took approximately 200,000 CPU-hours, spread over 3 weeks. Much larger problems are easy to construct --- either by considering larger systems of crossing equations or taking more derivatives. More crossing equations might arise from larger global symmetry groups, from considering four-point functions of operators with spin, from mixing different types of operators, or from a combination of these effects. In addition, more derivatives can lead to higher-precision results, so it is usually desirable to make $\Lambda$ as large as possible.

There are strong hints (from kinks and other features in numerical bootstrap bounds) that a more powerful semidefinite program solver could yield high-precision results for several important theories. With more a powerful solver, new CFTs could potentially become accessible as well. Furthermore, a more powerful solver could yield stronger general constraints on the space of CFTs, for example by improving and combining the bounds of \cite{Dymarsky:2017xzb,Dymarsky:2017yzx} on currents and stress tensors in 3d, and enabling generalizations to other spacetime dimensions and symmetry groups.

For these reasons, we have enhanced the semidefinite program solver SDPB to improve performance and scalability. The main changes are the replacement of OpenMP with MPI, which allows computations to scale across hundreds of cores on multiple nodes, and the replacement of MPACK \cite{mpack} with the Elemental library \cite{poulson2013elemental} for arbitrary precision distributed linear algebra.

The purpose of this note is to describe some technical details of the changes we have implemented and present performance comparisons between the old version of SDPB (1.0) and the new version (2.0). We do not give a detailed description of the algorithm, since it is the same as in version 1.0 \cite{Simmons-Duffin:2015qma}. We often refer to \cite{Simmons-Duffin:2015qma} for notation and definitions.

The structure of this note is as follows. In
section~\ref{sec:methods}, we describe our parallelization strategy
using MPI and a modified version of the Elemental library. In
section~\ref{sec:comparisons}, we describe changes relative to version
1.0 and verification of correctness. In section~\ref{sec:benchmarks},
we show benchmarks comparing the two versions of SDPB on two different
problems. We conclude in section~\ref{sec:futuredirections} with
discussion and future directions.

\subsection*{Code Availability}

SDPB is available from the repository at \url{https://github.com/davidsd/sdpb}.
The figures in this paper were created with version 2.1.3, which has
the Git commit hash
\begin{lyxcode}
6b1a5827c7b32481d7ab92fce1935c2f774a4d05
\end{lyxcode}
SDPB is written in C++, uses MPI for parallelism, and depends on Boost
\cite{boost}, GMP \cite{gmp}, MPFR \cite{mpfr}, and a forked version
of Elemental \cite{poulson2013elemental}. The fork is available at
\url{https://gitlab.com/bootstrapcollaboration/elemental}. The version
associated with this paper has the Git commit hash
\begin{lyxcode}
b39d6f5584092c0d56a68d2fc48aa653f0027565
\end{lyxcode}

\section{Methods\label{sec:methods}}

\subsection{Elemental and MPI\label{sec:elemental}}

SDPB solves semidefinite problems using a variety of dense linear
algebra operations using arbitrary precision numbers
\cite{Simmons-Duffin:2015qma}.  The original implementation relied on
the MPACK library \cite{mpack} for dense linear algebra routines, the
GMP library \cite{gmp} for arbitrary precision numerics, and OpenMP to
distribute work to multiple cores. This approach allowed SDPB to scale
well up to dozens of cores on a single, shared memory
node.

To scale to clusters with hundreds of cores on multiple nodes, we
replaced MPACK with the Elemental library \cite{poulson2013elemental,petschow2013high}.
Elemental implements routines for dense linear algebra on distributed
nodes, communicating via MPI. Elemental includes support for arbitrary
precision numeric types via the MPFR library \cite{mpfr}.

MPFR is very careful about correct rounding and exact handling
of precision, but, in our tests, ran about half as fast as GMP. SDPB
does not require correct rounding or exact handling of precision.
For example, it is acceptable for the user to ask for 1000 bits of
precision but actually get 1024 bits.

So we forked the Elemental library and replaced the MPFR routines
with GMP counterparts. This was not entirely trivial, because GMP does not implement
all of the functions that MPFR implements, such as the natural logarithm function \texttt{log()}.
For our purposes, where the most complicated linear algebra routines
that we use are Cholesky decomposition and eigenvalue computation,
the limited special function support in GMP was sufficient.

\subsection{Parallelization strategy\label{sec:parallelizationstrategy}}

To profitably use Elemental and MPI and achieve efficient scaling, we
had to partly restructure SDPB so that different cores work
independently as much as possible without waiting for each other to
finish.  For example, we had to be careful when computing global quantities, i.e.\
quantities that depend simultaneously on information distributed
across all cores.  An example of a global quantity is the primal error
defined below in (\ref{eq:correctprimalerror}), which requires
computing a maximum across all blocks of the matrix $P$.  If
a step of the computation depends on a global quantity, then the
remaining steps cannot be started until all cores have finished both
computing their contribution to that quantity and communicating that
information to each other.

Thus, achieving
good performance with MPI required us to be more deliberate when computing
anything that requires cores to communicate with one another. Fortunately,
most of the operations work on blocks made up of independent groups
of matrices. For example, the positive semidefinite matrix $Y$ (one of the internal variables in SDPB) can
be written as a block-diagonal matrix
\begin{align}
Y\equiv\left(\begin{array}{cccc}
Y_{1} & 0 & \cdots & 0\\
0 & Y_{2} & \cdots & 0\\
\vdots & \vdots & \ddots & \vdots\\
0 & 0 & \cdots & Y_{2J}
\end{array}\right),
\label{eq:Ymatrix}
\end{align}
where each $Y_{i}$ is a submatrix. Each pair of matrices $Y_{2n-1}, Y_{2n}$ is associated to a single positivity constraint. Thus, the number of submatrices $2J$ is essentially equal to twice the number of combinations of spins $\ell$ and symmetry representations $R$ appearing in the crossing-symmetry equations (\ref{eq:crossingequation}). For example, for the 3d Ising model computation described in section~\ref{sec:benchmarks}, we have $2J\approx 230$. Each iteration of SDPB involves several basic linear algebra operations involving the matrices $Y_i$ and other internal variables. Most of the operations associated with different positivity constraints are independent from each other, and so can be performed on separate cores without synchronization. An exception is the computation of the matrix $Q$, discussed in section~\ref{sec:global}.

\subsubsection{Block timings\label{subsec:Block-Timings}}

Matrix blocks in SDPB can vary substantially in how much time it takes to operate on
them. This can be because the blocks vary dramatically in size, or because large parts of the matrices are identically zero. Working
with the number zero in GMP is faster than other numbers. For example,
adding zero to a number only requires copying the number, while adding
general arbitrary precision numbers takes dramatically longer. The
structure of zeros within a block can be quite complicated, making
it difficult to create a priori estimates of how expensive each block
is.

Because computation times are difficult to predict, we instead directly measure
the time for the operations for each block. For each block, it turns out to be sufficient
to measure the time to compute its contribution to
$Q=B^{T}L^{-T}L^{-1}B$ (see Equation 2.44 in \cite{Simmons-Duffin:2015qma}).
In detail, this is the sum of the time to perform
\begin{itemize}
\item The Cholesky decomposition $S_{pq}=L_{pq}L_{pq}^{T}$, where $S_{pq}={\rm Tr}\left(A_{p}X^{-1}A_{q}Y\right)$,
\item The matrix solve $L_{pq}^{-1}B$,
\item The matrix multiplication $B^{T}L_{pq}^{-T}L_{pq}^{-1}B$.
\end{itemize}
During the first iteration of the solver, many numbers are still zero, so the first
iteration is usually faster than subsequent iterations. We have observed
that timings become stable after the first step, so we collect timings
from the second step. For these timing runs, the blocks are distributed
in a simple round-robin manner to all of the cores, with a single
core usually having many blocks. We write those block timings to a
file. Then we restart the calculation and use those timings to allocate
blocks to cores.

If there is no block timings file, SDPB will automatically perform a
timing run and restart. When starting from a checkpoint, such as when
hot-starting \cite{Go:2019lke}, SDPB will reuse block timings.  So a
timing run only happens when starting a new calculation from scratch.
Block timings can also be manually copied from previous calculations with the
same structure.

\subsubsection{Load balancing}

The problem at this point is then how to distribute blocks among cores
and nodes. This turns into a variant of the well-known problem of
bin-packing \cite{binpacking}. We want the most expensive blocks
to get more cores allocated to them. However, we still want to keep
the work for a single block on a single node, since latency and bandwidth
between nodes is much worse than within a node.

The algorithm we use is a variant of Worst-Fit \cite{worstfit}. We
assume that the cluster that SDPB is running on has $N$
identical nodes with $c$ cores each, totaling
$C=N \times c$
cores. Starting with all of the individual block timings
$t_{{\rm block}}$, we add them up to get a total time
\begin{align}
t_{{\rm total}}\equiv\sum_{\rm block}t_{{\rm block}}.
\end{align}
Dividing this by $C$ gives us an
average time per core

\begin{align}
t_{{\rm average}}\equiv\frac{t_{{\rm total}}}{C}.
\end{align}
We sort the blocks by their timings $t_{{\rm block}}$ and start with
the most expensive blocks. If the timing of
a block is greater than the average block time $t_{{\rm block}}>t_{{\rm average}}$,
then that block will get $t_{{\rm block}}/t_{{\rm average}}$ cores,
rounded down.

A block is assigned to nodes that have the most cores left. If there
are no nodes left that have enough spare cores, then the block is
squeezed into the node with the most cores left. This is the part
that is like Worst-Fit, since we assign blocks to the nodes
with the most available cores, rather than nodes with the least available cores
that can still fit each block.

This assigns all blocks with $t_{{\rm block}} > t_{{\rm average}}$.  Starting
with the largest remaining blocks, blocks are collected until the
sum of their timings is greater than or equal to the average timing.
That collection of blocks is then assigned to a single core. The core
will be on the node with the most remaining cores left. This process
is continued until all blocks are assigned.

This process necessarily underprovisions blocks, leaving left-over
cores on some nodes. These left-over cores are assigned, one-by-one,
to the block or collection of blocks with the highest timings per core.
For example, consider a setup with a node with 1 large block with $t_{{\rm block}\,1}=5$
that has two cores assigned to it (so $c_{{\rm block\,1}}=2$). In addition, that node has a collection of 2 smaller
blocks with 
\begin{align}
\begin{array}{ccc}
t_{{\rm collection}} & = & t_{{\rm block\,2}}+t_{{\rm block\,3}}\\
t_{{\rm block\,2}} & = & 1.2\\
t_{{\rm block\,3}} & = & 0.8
\end{array}
\end{align}
assigned to a single core. A left over-core would be assigned to the
large block, because 
\begin{align}
\frac{t_{{\rm block\,1}}}{c_{{\rm block\,1}}}=\frac{5}{2}>\frac{t_{{\rm collection}}}{c_{{\rm collection}}}=\frac{1.2+0.8}{1}.
\end{align}
For a second left-over core, the respective timings are 5/3 and 2.
In that case, the collection is split in two, with each block getting
a single core. For a third left-over core, the respective timings
are 5/3, 1.2/1, and 0.8/1. So the third core is assigned to the large
block.

This is by no means the most optimal solution. For example, when the
block collections are split up, they are not rebalanced amongst all
the collections on that node. Also, when a large block is squeezed
into a node, the existing blocks and block collections do not give up
any cores to make the distribution more equitable.  In practice, we
have observed a maximum imbalance of a factor of 2 between the work
assigned to cores.  This arises when there are two similar sized
blocks, each assigned one core.  An extra core is assigned to one of
the blocks and not the other.  This effect diminishes with larger core
counts, as the effect of a single core is less decisive.  For the
benchmarks in section~\ref{sec:benchmarks}, we observed imbalance factors
ranging from 1.02 to 1.2.

\subsubsection{Global operations}
\label{sec:global}

There are relatively few cases where there is a truly global operation
involving all of these different blocks. One significant case is in computing and subsequently inverting the
matrix $Q$, implemented as a globally distributed Elemental matrix.
When creating $Q$, every block contributes to every every element
of $Q$. So we compute the individual contributions of each group
of cores to $Q$.

We then perform a global reduce and scatter to sum all of the
contributions and distribute the results to the distributed version of
$Q$.  However, we found that if we used the builtin implementation of the function
\texttt{MPI\_Reduce\_scatter()} \cite{message2012mpi}, the MPI library
would allocate extra copies of $Q$ on each core for sending,
receiving, and summing.  When running on many cores, these extra
copies of $Q$ dominate the use of memory.  For large problems, this
extra memory usage could make it impossible to run SDPB.

To work around this, we reimplemented \texttt{MPI\_Reduce\_scatter()}.  In
particular, we implemented the ring algorithm as found in OpenMPI
\cite{gabriel04:_open_mpi}.  This allowed us to tightly control the
amount of memory used for temporary objects.  This hand-rolled
implementation causes some degradation in performance for larger jobs
(>50 cores), but it has dramatically reduced the memory usage.

With the global reduce and scatter completed, we use Elemental routines to
compute the Cholesky decomposition of $Q$ and distribute the result to
all of the different groups of cores.

\subsubsection{Memory Optimization}
\label{sec:memory}

While these changes did improve performance and scalability, they came
at the cost of significantly larger memory use.
The memory to store individual blocks is well distributed among the
cores. So the memory dedicated to blocks is more or
less identical between SDPB 2.0 and SDPB 1.0. However, as discussed
in Section \ref{sec:global}, there is a global matrix $Q$
that gets contributions from every block. In SDPB 1.0, there
is only one copy of this matrix. In SDPB 2.0, there is
a copy of the local contribution to $Q$ on every core. These contributions
are then added up to make the global $Q$. While $Q$ itself is not the largest internal matrix that the solver uses, all of these copies of $Q$ end up requiring
significant amounts of memory.

To mitigate this, we have added an option \texttt{procGranularity}.
The main effect of this option is to store fewer copies of $Q$ split
up among the cores.  Using this option makes the memory usage nearly
equivalent to the OpenMP version, but also makes the MPI version (2.0)
as slow as or slower than the OpenMP version.

\section{Comparisons with version 1.0\label{sec:comparisons}}

\subsection{Change to primal error\label{sec:Other-improvements}}

In the original SDPB paper \cite{Simmons-Duffin:2015qma}, the primal
error is defined as
\begin{align}
\label{eq:correctprimalerror}
{\rm primalError} & \equiv \max\left\{ \left|p_{i}\right|,\left|P_{ij}\right|\right\} ,\nonumber\\
P & \equiv  \sum_{i}A_{i}x_{i}-X,\nonumber\\
p & \equiv  b-B^{T}x.
\end{align}
 It turns out that SDPB 1.0 instead implemented 
\begin{align}
{\rm primalError}\equiv\max\left\{ \left|P_{ij}\right|\right\},
\end{align}
i.e.\ leaving out the contribution from the primal error vector $p$.
By contrast, in SDPB 2.0 we have implemented (\ref{eq:correctprimalerror}).
 Fortunately, the impact of this change appears to be minimal. In most cases we have checked,
if SDPB 1.0 finds a primal-dual optimal or primal feasible solution, then the missing part
of ${\rm primalError}$, $\left|p_{i}\right|$, is extremely small, and consequently doesn't affect
the actual value of ${\rm primalError}$.
For cases where $\left|P_{i}\right|$ is not small, then it
does not matter if $\left|p_{i}\right|$ is small or not.

For the case where there is only a primal solution, we have found
that if $\max\left\{ \left|P_{i}\right|\right\} $ has converged,
then $\max\left\{ \left|p_{i}\right|\right\} $ will eventually converge.
Usually it converges earlier, but sometimes, especially if starting
from a previous checkpoint \cite{Go:2019lke}, it can converge
later.

As an additional optimization, we removed the ``Cholesky
stabilization'' method described in \cite{Simmons-Duffin:2015qma}. The purpose of this method was to decrease the condition number of $Q$ before inverting it by adding low-rank contributions (and subtracting off their effects later). This method increases the size of $Q$, slowing down the computation. After experience with SDPB, we recommend
that users simply increase precision if they encounter numerical stability issues. This also slows down the computation, but in practice less so than Cholesky stabilization. When initializing the Schur complement
solver, removing Cholesky stabilization also allowed us to use a Cholesky decomposition,
rather than an LU decomposition, of the $Q$ matrix. This optimization
has a small, but noticeable, effect on the total time.

\subsection{Correctness check}

To ensure that no errors were introduced when upgrading SDPB, we ran a number of problems with both the old and new versions.
One example problem is a feasibility search in the 3d Ising model. After two steps, for runs with
 200 decimal digits of accuracy; the typical difference was around the
180th digit. Some differences are expected, since the exact nature and order of
operations to compute quantities like eigenvalues and Cholesky decompositions
will be different between MPACK and Elemental. This comparison gives evidence that the basic linear algebra operations are the same between the two versions. The differences between numbers in the old and new versions increase with more iterations.
After 100 iterations, the exact values can be quite different in low decimal places. However, both versions do arrive at the same solution. For optimizations (as opposed to feasibility searches), where the solver is used to minimize or maximize some quantity by finding a primal-dual solution, both versions found the same result to high
accuracy (80 digits).

\section{Benchmarks\label{sec:benchmarks}}

To demonstrate improved performance and scalability, we have
benchmarked the new and original versions. All benchmarks were
performed on the Grace cluster at Yale. Each node in the Infiniband
networked cluster is a Lenovo nx360b, containing 28 cores across
two Intel E5-2660 v4 CPU's and a total of 247 GB of memory. The first
benchmark is the semidefinite program described in
\cite{Kos:2014bka,Simmons-Duffin:2015qma} that arises from studying
four-point functions of the $\sigma$ and $\epsilon$ operators in the
3d Ising model.  In the notation of the introduction, it involves $I_\mathrm{max}=5$ crossing equations, OPE
coefficient vectors of size $\dim(\vec\lambda_{R,\Delta,\ell})\in\{1,2\}$, maximum derivative order $\Lambda=43$, and a functional with $N=1265$ components.  The second,
larger, benchmark, described in \cite{Go:2019lke,O2Future}, arises
from studying four-point functions of the leading-dimension charge-0,
1, and 2 scalar operators in the $O(2)$ model.  It involves $I_\mathrm{max}=22$ crossing
equations, OPE coefficient vectors of size $\dim(\vec\lambda_{R,\Delta,\ell})\in\{1,2,3\}$,
maximum derivative order $\Lambda=35$, and a functional with $N=3762$ components.  Both benchmarks were run with 1216 bits of precision.
We measure the time to compute the second iteration, which, as noted
in Section \ref{subsec:Block-Timings}, we have found to be a stable
and accurate measurement of the overall performance of the solver.

\begin{figure}
\centering{}\includegraphics[width=0.6\paperwidth]{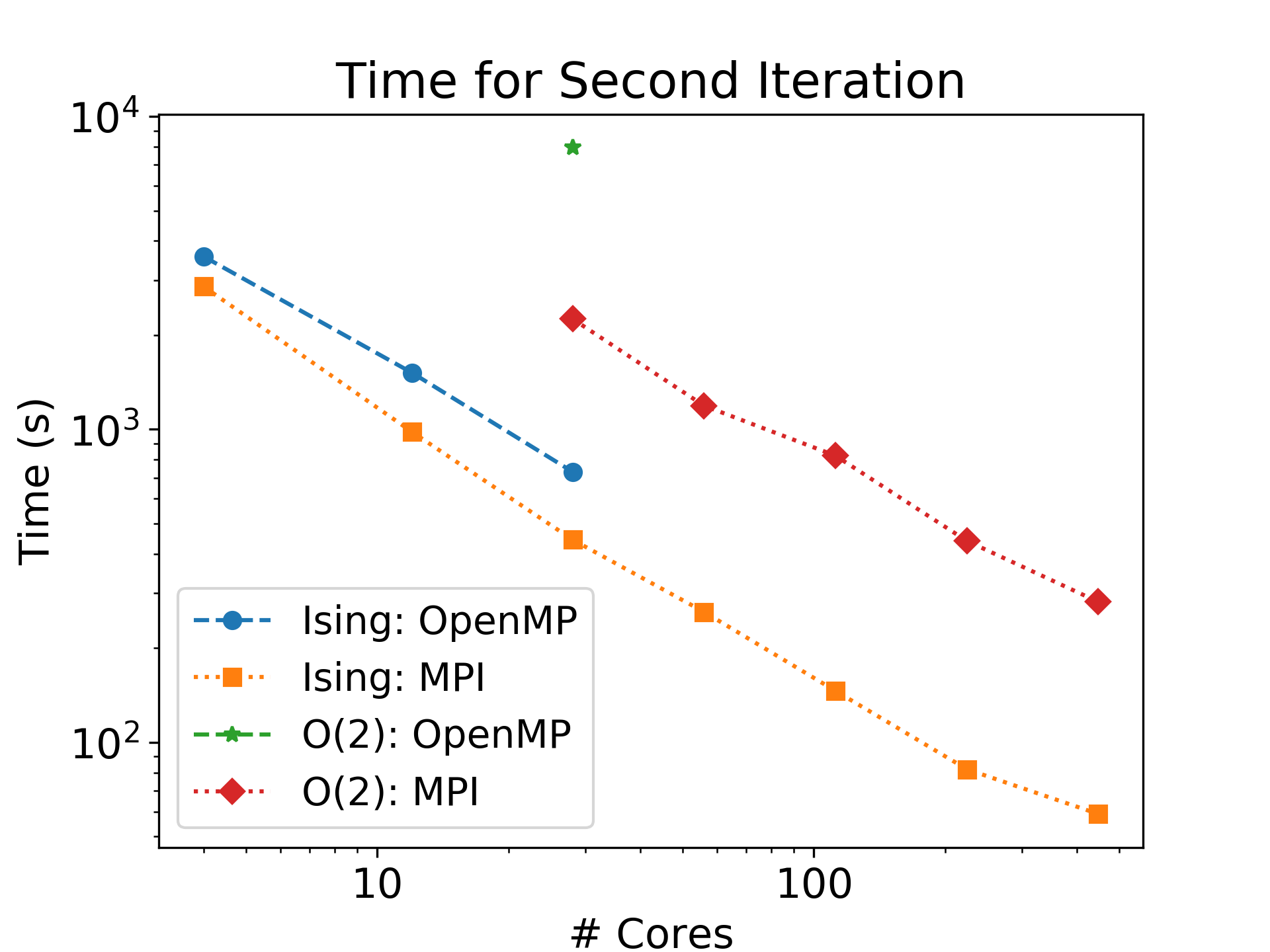}\caption{\label{fig:Scaling-Time}Time to compute the second iteration of an
Ising model with 117 blocks (as in \cite{Simmons-Duffin:2015qma})
and an O(2) model with 261 blocks on the Yale Grace cluster. SDPB 1.0
is limited to the number of cores on a single node (28), while SDPB 2.0
can scale across multiple nodes.}
\end{figure}

Figure \ref{fig:Scaling-Time} and Table \ref{tab:Scaling-Time}
demonstrate that SDPB 2.0 is always faster, with the difference
becoming more dramatic at higher core counts. At low core counts, the
remaining difference is largely due to using a Cholesky decomposition
rather than an LU decomposition (see Section
\ref{sec:Other-improvements}).

The scaling of SDPB 2.0 is quite good, and gets better for larger
problems.  The performance does drop off at high core counts, most likely due
to our memory-optimized version of \texttt{MPI\_Reduce\_scatter()} (Section \ref{sec:global}).

\begin{figure}
\centering{}\includegraphics[width=0.6\paperwidth]{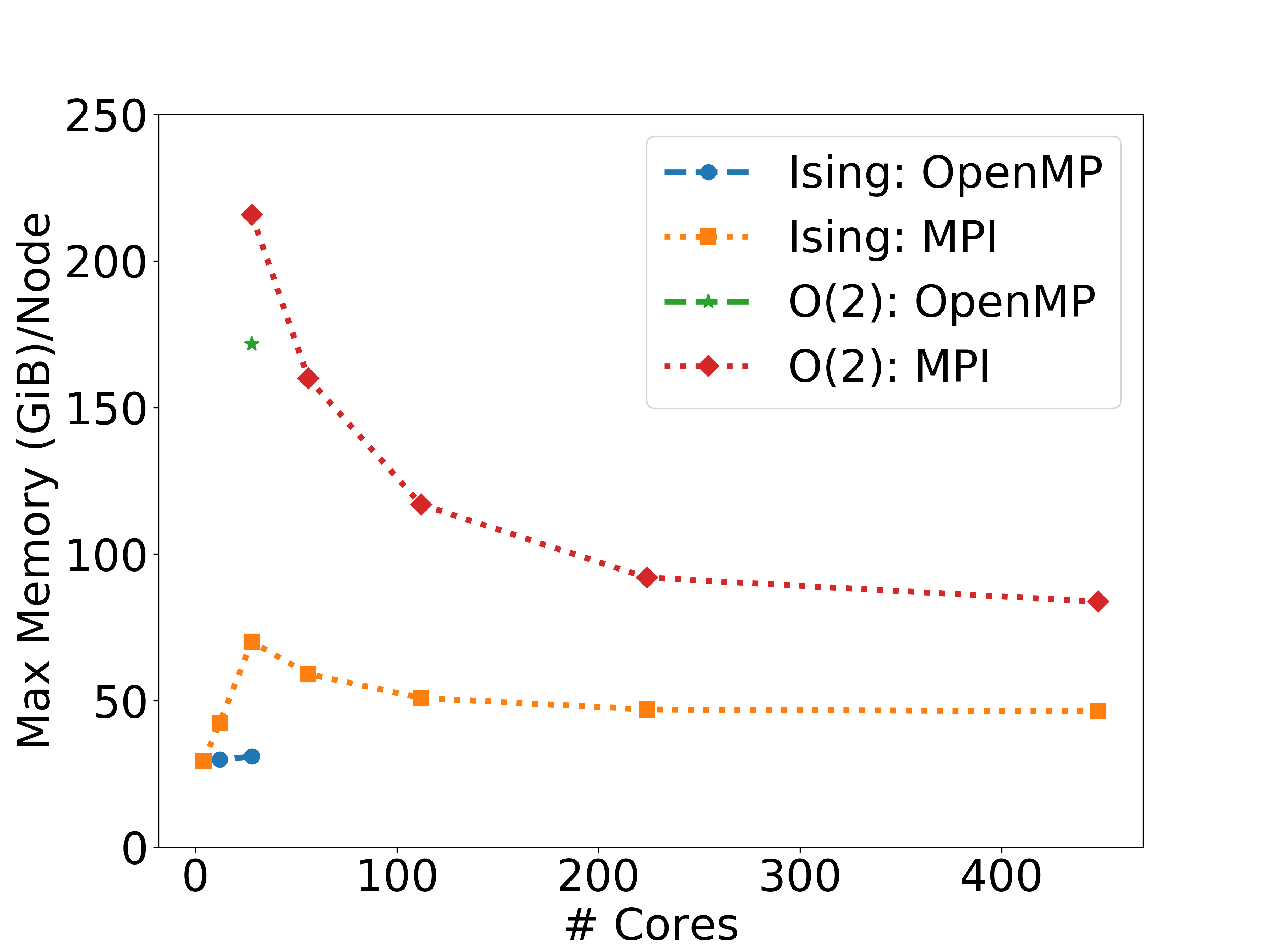}\caption{\label{fig:Scaling-Memory}Maximum memory used on a node for the computation
in Figure \ref{fig:Scaling-Time}.}
\end{figure}

Figure \ref{fig:Scaling-Memory} and Table \ref{tab:Scaling-Memory}
show the maximum memory required on a node. SDPB 1.0's memory use is
relatively constant.  SDPB 2.0 uses significantly more memory,
continually increasing with the number of cores per node.  At higher
node counts, the memory use per node stabilizes.  The computation was
able to fit on the available nodes, so we did not use the memory
optimization flag detailed in section~\ref{sec:memory}.

\begin{table}[ht]
\centering{}%
\begin{tabular}{cc}
3d\ Ising & $O(2)$\tabularnewline
\begin{tabular}{|c|c|c|}
\hline 
\#Cores & OpenMP & MPI\tabularnewline
\hline 
\hline 
4 & 3568 & 2861\tabularnewline
\hline 
12 & 1517 & 982\tabularnewline
\hline 
28 & 732 & 444\tabularnewline
\hline 
56 & & 261\tabularnewline
\hline 
112 & & 146\tabularnewline
\hline 
224 & & 82\tabularnewline
\hline 
448 & & 59\tabularnewline
\hline 
\end{tabular} & %
\begin{tabular}{|c|c|c|}
\hline 
\#Cores & OpenMP & MPI\tabularnewline
\hline 
\hline 
& &\tabularnewline
\hline 
& &\tabularnewline
\hline 
28 & 7950 & 2254\tabularnewline
\hline 
56 & & 1190\tabularnewline
\hline 
112 & & 824\tabularnewline
\hline 
224 & & 440\tabularnewline
\hline 
448 & & 281\tabularnewline
\hline 
\end{tabular}\tabularnewline
\end{tabular}\caption{\label{tab:Scaling-Time}Time (s) to compute iterations in Figure \ref{fig:Scaling-Time} }
\end{table}

\begin{table}[ht]
\centering{}%
\begin{tabular}{cc}
3d\ Ising & $O(2)$\tabularnewline
\begin{tabular}{|c|c|c|}
\hline 
\#Cores & OpenMP & MPI\tabularnewline
\hline 
\hline 
4 & 29.4 & 29.3\tabularnewline
\hline 
12 & 29.9 & 42.3\tabularnewline
\hline 
28 & 31.0 & 70.1\tabularnewline
\hline 
56 & & 59.1\tabularnewline
\hline 
112 & & 50.9\tabularnewline
\hline 
224 & & 47.0\tabularnewline
\hline 
448 & & 46.4\tabularnewline
\hline 
\end{tabular} & %
\begin{tabular}{|c|c|c|}
\hline 
\#Cores & OpenMP & MPI\tabularnewline
\hline 
\hline 
& &\tabularnewline
\hline 
& &\tabularnewline
\hline 
28 & 172 & 216\tabularnewline
\hline 
56 & & 160\tabularnewline
\hline 
112 & & 117\tabularnewline
\hline 
224 & & 92.0\tabularnewline
\hline 
448 & & 83.8\tabularnewline
\hline 
\end{tabular}\tabularnewline
\end{tabular}\caption{\label{tab:Scaling-Memory}Maximum memory (GiB) used on a node for the computation
in Figure \ref{fig:Scaling-Time} }
\end{table}

\section{Future directions\label{sec:futuredirections}}

We have modified SDPB to use MPI and the Elemental library, and
demonstrated that it scales well up to hundreds of cores. There are
several interesting directions for future optimizations and
improvement. Perhaps the most urgent is to decrease the precision
required for numerical bootstrap computations. In practice,
large-scale computations require approximately 1000 bits of precision
--- roughly 20 times what is used in IEEE double precision
arithmetic (53 bits for the mantissa). The exact reasons why such precision is
necessary are currently unknown, but we do notice that matrices internal to SDPB
develop very large condition numbers during computations. Inverting or
Cholesky-decomposing these matrices causes instabilities if the
precision of the arithmetic is too low. For example, the blocks of the
``Schur complement'' matrix $S_{pq}$ (see section
\ref{subsec:Block-Timings}) can develop condition numbers of size
$10^{400}$, while the condition number of $Q$ can reach
$10^{100}$. This suggests that a clever change of basis for {\it
  both\/} the primal and the dual problems might be necessary to
mitigate precision issues.

Even with a clever change of basis, we still expect it would be extremely
difficult to use machine-precision arithmetic in numerical bootstrap
computations. This would require a far more detailed understanding of
solutions to crossing symmetry than we currently possess.\footnote{See
  \cite{Afkhami-Jeddi:2019zci} for an example of enormous gains that
  can come from a detailed understanding of solutions to crossing
  symmetry, in the context of the modular bootstrap
  \cite{Hellerman:2009bu}.} Historically, high precision has been a
powerful tool for circumventing these issues and simply getting things
done. One possible future direction is to dynamically change precision
based on the size of the problem and the needs of the solver, with less delicate parts of the
computation running at lower precision. Precision could be increased
slowly over multiple iterations, so that high precision is only used
near the end of the computation when condition numbers are
largest. (This particular optimization may be less useful in
conjunction with hot-starting \cite{Go:2019lke,O2Future}.) More
optimistically, perhaps high condition numbers and numerical
instabilities can be avoided using a different semidefinite
programming algorithm, or alternative optimization methods.

Another challenge for SDPB is to scale to even larger machines and
larger problems.  As one example, our implementation of
\texttt{MPI\_Reduce\_scatter()} could certainly be optimized.  Comparing our
hand-rolled version with the OpenMPI implementation, there is no
obvious reason that our implementation is slower at high core counts.
This points to a general need to revisit other communication and
synchronization procedures in MPI and optimize them for our use case.

Modern high performance computing clusters often include GPUs, which
are in principle well-suited for highly parallel linear algebra
operations. We could leverage existing libraries for high-precision linear
algebra on GPUs \cite{10.1007/978-3-319-42432-3_29,CUMP}.\footnote{We thank Luca Iliesiu and Rajeev Erramilli for discussion on this point.}

SDPB 1.0 demonstrated the usefulness of a robust optimization tool specialized for numerical bootstrap applications. We hope that SDPB 2.0 will demonstrate the usefulness of scaling such a specialized tool to large problems. In the meantime, it is worthwhile to explore other optimization methods that might prove more efficient in the asymptotic future. The design of SDPB 1.0 was based on the interior point method implemented in SDPA \cite{SDPA,SDPA,SDPAGMP}. Perhaps other SDP algorithms like the multiplicative weight method \cite{Arora:2016:CPA:2906142.2837020} suggested in \cite{Bao:2018dzk} could be more efficient, or circumvent the precision problems of interior point methods. Perhaps the primal-dual simplex method of \cite{El-Showk:2014dwa,Paulos:2014vya} can be adapted to work with matrix positivity conditions. The implementation of \cite{El-Showk:2014dwa} dealt inefficiently with double-roots of extremal functionals. Perhaps it can be modified to allow roots to move continuously, incorporating ideas from \cite{El-Showk:2016mxr,Afkhami-Jeddi:2019zci}.\footnote{We thank Tom Hartman for discussion on this point.} While these are superficially engineering problems, their solutions could have deep implications and broad consequences for our understanding of strongly-coupled field theories.

\section*{Acknowledgements}

We thank Ning Bao, Connor Behan, Shai Chester, Rajeev Erramilli, Tom Hartman, Luca Iliesiu, Filip Kos, Junyu Liu, Petr Kravchuk, David Poland, Ning Su, and Alessandro Vichi for discussions. We also thank the participants in the Simons Bootstrap Collaboration kickoff meeting at Yale in 2016 and the bootstrap ``hackathon'' at Caltech in summer 2018. DSD and WL are supported by Simons Foundation grant 488657 (Simons Collaboration on the Nonperturbative Bootstrap). DSD also acknowledges support from a Sloan Research Fellowship, and a DOE Early Career Award under grant No.\ DE-SC0019085.

\bibliographystyle{JHEP}
\bibliography{reference}

\end{document}